\documentstyle[preprint,aps]{revtex}
\begin{document}
\draft
\title{Electromagnetic Corrections to $\pi\pi$ Scattering Lengths:
Some \\ Lessons for the Implementation of Meson Exchange Models}
\author{Kim Maltman}
\address{Department of Mathematics and Statistics, York University,
4700 Keele St., \\ North York, Ontario, Canada M3J 1P3}
\author{Carl E. Wolfe}
\address{Department of Physics and Astronomy, York University,
4700 Keele St., \\ North York, Ontario, Canada M3J 1P3}
\date{\today}
\maketitle
\begin{abstract}
The leading (in chiral order) electromagnetic corrections to 
s-wave $\pi\pi$ scattering lengths are computed using chiral
perturbation theory (ChPT).  It is shown that contributions
associated, not with one photon exchange, but rather with
contact terms in the effective electromagnetic Lagrangian,
dominate previously computed radiative corrections (which are
higher order in the chiral expansion). These corrections bring
experimental values into closer agreement with the results of
ChPT to one-loop order.  It is also pointed out that 
standard implementations of electromagnetism in the meson-exchange
model framework omit such contact terms and that this omission,
combined with experimental input, would lead to the erroneous
conclusion that $\pi\pi\rightarrow\pi\pi$ exhibited very
large strong isospin breaking.  Implications for standard
``electromagnetic subtraction'' proceedures, and for the
construction of meson-exchange models in general, are also discussed.
\end{abstract}
\def\leff{${\cal L}_{eff}\ $}
\def\leffpunc{${\cal L}_{eff}$}
\def\ogame{$1\gamma E\ $}
\def\oqz{${\cal O}(q^0)\ $}
\def\oq2{${\cal O}(q^2)$\ }
Over the past decade or so, there has been considerable debate
between proponents of meson-exchange models and those models/methods
involving explicit quark and gluon degrees of freedom, over which
approach to treating
strongly interacting few-body systems is the correct one.
In the most general context, however, this debate is, in fact,
meaningless.  Indeed, one is free to restrict one's attention to a
set of dynamical, low-lying hadronic degrees of freedom and write a
low-energy effective Lagrangian, \leffpunc , which involves explicitly
only these degrees of freedom.  So long as this Lagrangian is
constructed in such a way as to be the {\it most general} such
Lagrangian possessing all the exact symmetries of QCD and realizing
the approximate chiral symmetries of QCD in the same way
they are realized in QCD, it will, of course, be identical
to QCD in its consequences for any processes involving only the
explicitly considered degrees of freedom --- provided, that is, the
unknown coefficients accompanying the terms in \leff allowed by the
symmetry arguments (called ``low-energy constants'', or LEC's)
are given the values they would have in QCD.  One can then view
typically implemented meson-exchange models as truncations of the
most general such low-energy effective theory -- truncations in which
certain types of terms are omitted and in which the values of the LEC's
accompanying other terms are not computed from QCD (usually an impossible
task, at present), but rather are left free, to be determined
phenomenologically.  These truncations may, of course, do damage
to the underlying theory.  In the untruncated version, however, the
effective theory fully incorporates QCD, despite having no explicit
quark and gluon degrees of freedom.  Their effect is present,
but buried in the precise values of the LEC's which describe
the non-renormalizable low-energy effective theory.

In order to set the context for the above remarks, it is useful to
remember that there exists a general framework for constructing,
for a given restricted set of hadronic degrees of freedom, the
most general low-energy effective theory compatible with the
symmetries and approximate symmetries of 
QCD\onlinecite{cwz,ccwz}.  A crucial feature of the resulting
effective Lagrangians is the presence of contact interactions
involving, in general, arbitrary numbers of hadronic fields.
These contact interactions are present because, in constructing
\leffpunc , one has effectively integrated out high frequency components.
Certain n-point Green functions generated by explicit quark and
gluon loops in QCD then generate, from their high-momentum (short
distance) components, effective interactions which appear
point-like from the perspective of the effective theory.  Such
contact terms are almost always omitted in formulating 
meson-exchange models.  However, to the extent that they correspond
to terms allowed by the symmetries of QCD, they are a necessary
consequence of QCD, and must be present in \leffpunc , regardless of
whether or not this complicates the phenomenological task of
fixing the values of the full set of unknown LEC's
relevant to a given process.  Expanding the set of hadronic
degrees of freedom included in \leff may alter the values of the
LEC's (and introduce new LEC's, as well), but for any finite
truncation of the hadronic degrees of freedom, such contact
terms will necessarily be present.
Moreover, as we will show in this letter using the example
of electromagnetic contributions to
$\pi\pi$ scattering, omission of these contact terms
can lead to serious numerical errors in the treatment of
the physics of 
a given problem.

Let us now illustrate the above comments by showing explicitly
how the effects of electromagnetism (EM) in hadronic systems
would be incorporated in the effective Lagrangian framework.
The discussion will also serve to fix our notation.  
We adhere throughout to the general framework for constructing
\leff developed in in Refs.~\onlinecite{cwz,ccwz,weinberg79,gl85}
(see also Refs.~\onlinecite{eckerrev,meissnerrev} for excellent
recent reviews).

If one imagines an effective theory involving, say, the pseudoscalar
and vector mesons and octet baryons, then, as is well-known,
one may make field choices for the particles such that,
with $B$ the standard octet baryon matrix, $S_{\mu\nu}$ and
$V_{\mu\nu}$ the singlet and flavor-octet matrix for the
vector mesons in the tensor field 
representation\onlinecite{gl84,ecker89}, $\pi\equiv \lambda^a\pi^a$
the standard octet pseudoscalar matrix, $U=\exp (i\pi /F)$, and
$u=\exp (i\pi /2F)$ (with F a parameter which turns out to be
the $\pi$ decay constant in the chiral limit), the
fields transform as
\begin{eqnarray}
&&B\rightarrow h\, B\, h^\dagger \label{one}\\
&&V_{\mu\nu}\rightarrow h\, V_{\mu\nu}\, h^\dagger\label{two} \\
&&S_{\mu\nu}\rightarrow S_{\mu\nu} \label{three} \\
&&U\rightarrow R\, U\, L^\dagger \label{four} 
\end{eqnarray}
under the chiral group $SU(3)_L\times SU(3)_R$\onlinecite{cwz,ccwz,georgi}.
Here, $h$ is defined (as a function of $\pi^a$ and the left and right
chiral transformation matrices, $L$ and $R$) via
\begin{equation}
u\rightarrow R\, u\, h^\dagger = h\, u\, L^\dagger\label{five}
\end{equation}
and reduces to the ordinary $SU(3)_V$ transformation matrix $V=L=R$
when the chiral transformation lies in the vector subgroup $SU(3)_V$.
In the chiral limit (no EM and $m_u=m_d=m_s=0$), \leff consists
of all terms involving the field variables and their derivatives
which are invariant under the full chiral group (as well as under
C, P, T and Lorentz transformations).  The construction of \leff
is, as usual, greatly simplified by introducing the covariant
derivatives of the various fields, which, by construction, transform
in the same way as the original fields, e.g.
\begin{eqnarray}
&&D_\mu U\ \equiv\ \partial_\mu U -\ i\, r_\mu\, U\ +\ i\, U\,\ell_\mu
\rightarrow R(D_\mu U)L^\dagger\nonumber \\
&&D_\mu B \ \equiv\ \partial_\mu B\ +\ [\hat{V}_\mu ,B]\rightarrow
h(D_\mu B)h^\dagger\ etc.,\label{six}
\end{eqnarray}
where $\hat{V}_\mu\equiv {\frac{1}{2}}\left[ u^\dagger (\partial_\mu
-ir_\mu )u\ +\ u(\partial_\mu -i\ell_\mu )u^\dagger\right]$
with $\hat{V}_\mu$ 
transforming as $\hat{V}_\mu\rightarrow h\hat{V}_\mu h^\dagger
\ +\ h\partial_\mu h^\dagger$, where $r_\mu$ and $\ell_\mu$ are
the external right and left-handed sources which, for example,
allow one to treat explicit couplings to photons and W bosons.
(Since $r_\mu =v_\mu\ +\ a_\mu$, $\ell_\mu =v_\mu\ -\ a_\mu$, with
$v_\mu$ and $a_\mu$ the external vector and axial vector spurces, the
choice $v_\mu =\, -eQA_\mu$, for example, with $Q$ the quark
charge matrix, generates the explict couplings to photons.)
The external field tensors, $L_{\mu\nu}$, $R_{\mu\nu}$, which
transform as $L_{\mu\nu}\rightarrow L\,L_{\mu\nu}\, L^\dagger$,  
$R_{\mu\nu}\rightarrow R\,R_{\mu\nu}\, R^\dagger$, and their
covariant derivatives can also occur in \leffpunc .  To facilitate 
construction of the couplings of the pseudoscalars to the baryons
and vector mesons, it is also convenient to introduce, in 
addition to $\hat{V}_\mu$ above, the combination
$\hat{A}_\mu\ \equiv\ {\frac{i}{2}}\left[ u^\dagger (\partial_\mu
-ir_\mu )u\ -\ u(\partial_\mu -i\ell_\mu )u^\dagger\right]$
which transforms as $\hat{A}_\mu\rightarrow h\ \hat{A}_\mu\, h^\dagger$.
Similarly, the couplings of baryons and vector mesons to the
external field tensors are simplified by employing
\begin{equation}
F_\pm^{\mu\nu}
\ \equiv\ u\, L^{\mu\nu}\, u^\dagger\ \pm\ 
u^\dagger\, R^{\mu\nu}\, u\label{nine}
\end{equation}
which transform as $F_\pm^{\mu\nu}\rightarrow h\, F_\pm^{\mu\nu}\, h^\dagger$.
The basic ingredients for constructing \leff are then traces of products
of the building blocks above and their covariant derivatives, where the
terms are ordered inside a given trace in such a way as to
produce manifest invariance.

EM and quark masses, which explicitly break the chiral invariance,
are also easily incorporated by noting that the chiral symmetry
breaking piece of ${\cal L}_{QCD}$, ${\cal L}_{\chi SB}$,
\begin{equation}
{\cal L}_{\chi SB}\ =\ -\bar{\psi_L}M_{LR}\psi_R -\  
\bar{\psi_R}M_{RL}\psi_L\ +\ eA_\mu\left( 
\bar{\psi_L}Q_{LL}\gamma^\mu\psi_L
\ +\ \bar{\psi_R}Q_{RR}\gamma^\mu\psi_R\right)\label{ten}
\end{equation}
(where $M_{LR}=M_{RL}=M$ is the quark mass matrix, and
$Q_{LL}=Q_{RR}=Q$ is the quark charge matrix) would be
invariant if $M_{LR}$ {\it etc.} were thought of as spurions
transforming as indicated by the subscripts, {\it i.e.},
$M_{LR}\rightarrow L\, M_{LR}\, R^\dagger$, {\it etc.}.
The chiral symmetry breaking terms in \leff are then generated by
including explicit factors of $M$, $Q$ to produce terms which are
invariant under the action of the joint field and spurion
transformation rules above.  (For convenience of use in the
baryon and vector meson sectors, one may use, in place of $M$,
$Q$, $M_\pm$ and $Q_\pm$, where
\begin{eqnarray}
M_\pm\ =\ {\frac{1}{2}}\left( u^\dagger\, M_{RL}\, u^\dagger\ \pm \ 
u\, M_{LR}\, u\right)\nonumber \\
Q_\pm\ =\ {\frac{1}{2}}\left( u^\dagger\, Q_{RR}\, u\ \pm \ 
u\, Q_{LL}\, u^\dagger\right)\label{twelve}
\end{eqnarray}
which transform as $M_\pm\rightarrow h\, M_\pm\, h^\dagger$,
$Q_\pm\rightarrow h\, Q_\pm\, h^\dagger$.)

Let us now employ the above machinery to incorporate EM in \leffpunc .
The first class of contributions is familiar from the conventional
treatments of EM in the meson-exchange framework.  It consists
of those graphs involving explicit photons, either via one-photon
exchange (\ogame ) or through radiative corrections ({\it i.e.}
graphs with photon loops).  The photon couplings
of the mesons and baryons are generated by the presence of the
external vector field $v_\mu$, which includes the photon, in the
covariant derivatives of the baryon and meson fields, and
also by the possibility of couplings involving the external
field tensors, $L^{\mu\nu}$, $R^{\mu\nu}$ (or, equivalently,
$F_\pm^{\mu\nu}$).  An example of the latter would 
be\onlinecite{ecker89}
\begin{equation}
{\cal L}_{\gamma V}
\ =\ {\frac{F_V}{2\sqrt{2}}}\, {\rm Tr}
\left[ V_{\mu\nu}F_+^{\mu\nu}\right]\label{thirteen}
\end{equation}
a P,C,T, Lorentz and chiral invariant term which, at zeroth
order in the pseudoscalar fields, couples the $\rho^0$ and
$\omega_8$ to the photon.  The LEC $F_V$ is then the vector
meson decay constant in the chiral limit.  As usual, its value
is not fixed by the symmetry arguments used in obtaining
${\cal L}_{\gamma V}$, which arguments tell us only that a term of the 
form ${\cal L}_{\gamma V}$ must be present in \leffpunc .
The value of $F_V$ is to be fixed from experiment, or
calculated explicitly in QCD.  Note that ${\cal L}_{\gamma V}$,
upon expanding the exponentials in $u$, $u^\dagger$ appearing
in $F_+^{\mu\nu}$, also involves contact couplings of a
photon to a single vector meson plus arbitrary numbers of
pseudoscalars.

The explicit photon-exchange and radiative correction contributions
to a given process do not, however, exhaust the set of EM effects.
As usual, because we have an effective low-energy theory,
certain high frequency photon-exchange and -loop contributions
are effectively frozen out and represented by EM contact
terms in \leffpunc , {\it i.e.} terms which involve two powers of
the quark charge matrix, but which involve no explicit photons.
It is easy to see, by construction, that such terms must appear
in \leffpunc .  If we restrict our attention to terms which are
momentum-independent then, for example, for the pseudoscalar
sector, there is a unique such term,
\begin{equation}
{\cal L}^{(0)}_{\pi ,EM}
\ =\ c_\pi F^2 {\rm Tr}\left[ 
U_{RL}Q_{LL}U^\dagger_{LR}Q_{RR}\right]\label{fourteen}
\end{equation}
where the insertion of the factor $F^2$ is conventional, the
superscript indicates the chiral order (no derivatives and
no powers of $M$, in this case), and the chiral transformation
labels have been included as subscripts so one can see the
manifest chiral invariance of ${\cal L}^{(0)}_{\pi ,EM}$.
This term generates contributions to the $\pi^\pm$, $K^\pm$
masses and reproduces (in a very efficient manner) the
current algebra relations for the pseudoscalar EM 
self-energies\onlinecite{dashen}.

Similarly, if one wished to consider EM contributions to
$\rho$-$\omega$ mixing, one would have not only the usually
considered contributions, associated with an intermediate
one-photon state, but also contributions from terms in
\leff of the form
\begin{eqnarray}
{\cal L}^{(0)}_{V,EM}\ &=&\ c_V^{(1)} {\rm Tr}\left[ V_{\mu\nu}V^{\mu\nu}
Q_+^2\right]\ +\ c_V^{(2)} {\rm Tr}\left[ V_{\mu\nu}Q_+^2\right]
S^{\mu\nu}\ + \ c_V^{(3)} {\rm Tr}\left[ V_{\mu\nu}Q_+V^{\mu\nu}
Q_+\right]\ +\nonumber \\
&&c_V^{(4)} {\rm Tr}\left[ V_{\mu\nu}Q_+\right]
{\rm Tr}\left[ V^{\mu\nu}Q_+\right]\ +\ \cdots\label{fifteen}
\end{eqnarray}
where $+\cdots$ refers to other zeroth order EM terms which
do not contribute to $\rho$-$\omega$ mixing and which, hence,
have not been written down explicitly.

EM zeroth order contact terms also exist for the 
baryons\onlinecite{cg,lebed}:
\begin{eqnarray}
&&{\cal L}^{(0)}_{B ,EM}\ =\ c_B^{(1)} {\rm Tr}\left[ \bar{B}Q_+^2B
\right]\ +\ c_B^{(2)} {\rm Tr}\left[ \bar{B} BQ_+^2\right]\ +\ 
c_B^{(3)} {\rm TR}\left[ Q_+\bar{B} Q_+B\right] \nonumber \\
&&\qquad\qquad +\ c_B^{(4)} {\rm Tr}\left[ \bar{B}Q_+\right]
{\rm Tr}\left[ BQ_+\right]\ +\ c_B^{(5)} {\rm Tr}\left[
Q_+^2\right]{\rm Tr}\left[ \bar{B} B\right]\label{sixteen}
\end{eqnarray}
(not all of these terms are linearly independent\onlinecite{lebed}, 
but we show
the full set of manifestly chiral invariant terms that can be
constructed, for completeness).

Van Kolck, Friar and Goldman have, similarly, demonstrated
the presence of such EM contact terms for $\pi N$
interactions\onlinecite{vkfg}.

All of the above EM terms are {\it necessarily} present in
\leffpunc .  Any attempt to evaluate EM effects ignoring them will
result in incorrect results, {\it i.e.}, results incompatible
with the constraints of QCD.

Let us now turn to the case of EM contributions to $\pi\pi$
scattering in order to demonstrate that the contact
(``invisible photon'') contributions can, in fact,
be numerically crucial.  We will make non-trivial use of
the chiral counting, or low-energy expansion, in which
contributions to physical observables are organized in
a joint series in momenta and quark masses.  Because
of the leading chiral result that $m_\pi^2 \equiv q_\pi^2$ is
proportional to $(m_u +m_d)$, it is necessary to count
$M$ as ${\cal O}(q^2)$.  The terminology ``zeroth order''
then means no momenta and no quark masses, ``second
order'' two powers of momenta or one power of quark
masses, etc..  Using this counting, and the fact that
the charged meson-photon couplings are ${\cal O}(q)$
and the photon propagator ${\cal O}(q^{-2})$, the
\ogame graphs in $\pi\pi$ scattering are seen to be
${\cal O}(q^0)$.  The leading strong contributions
to $\pi\pi$ scattering are produced by the lowest
order part of the effective strong Lagrangian
\begin{equation}
{\cal L}_{st}^{(2)}\ =\ {\frac{F^2}{4}}
{\rm Tr}\left[ D_\mu U D^\mu U^\dagger \right]\ +\ 
B_0{\frac {F^2}{2}}{\rm Tr}\left[ MU^\dagger \ +\ UM\right]
. \label{seventeen}
\end{equation}
The LEC, $F$, is the $\pi$ decay constant in the chiral
limit, while the second LEC, $B_0$, determines the quark condensate
in the chiral limit via $<0\vert \bar{q}q\vert 0>=
-B_0F^2$.  ${\cal L}_{st}^{(2)}$ gives rise, at
leading order, to the Weinberg results for the $\pi\pi$
scattering lengths and slope parameters\onlinecite{weinberg66}.
Using the Weinberg counting argument\onlinecite{weinberg79},
one easily sees that the radiative corrections to 
the tree-level strong scattering graphs are \oq2
in the chiral counting.

It should be noted that, despite the arguments above, which
show the necessity of EM contributions not involving
explicit photons in any low-energy effective hadronic theory,
standard implementations in the
meson-exchange-model framework typically include only
contributions associated with the presence of explicit
photons.  Frequently, morover, it is only the \ogame
contributions which are considered, as, for example,
in the ``EM subtraction'' performed on NN scattering
data to determine the strong charge-independence-
and charge-symmetry-breaking scattering length 
differences $a_{pp}-a_{np}$ and $a_{pp}-a_{nn}$.
Certain other calculations do include EM radiative corrections,
as in the case of Morrison's treatment of EM corrections
to $\pi NN$ couplings\onlinecite{morrison} or,
particularly pertinent to the case at hand, Roig and Swift's
treatment of EM corrections to the $\pi\pi$
s-wave scattering lengths\onlinecite{roig76} (in which
\ogame contributions, including the effect of the $\pi$
EM form factor, as well as the full set of radiative
and bremstrahlung corrections, were taken into account).
However, apart from corrections to $\pi\pi$ 
scattering\onlinecite{roig76} and pionium
decay\onlinecite{moor95}
associated with the use of different masses for the
neutral and charged pions (since this splitting is
known to be essentially pure EM\onlinecite{gl82},
this amounts to taking into account the effect of
those parts of the EM contact interaction of
Eqn.~(\ref{fourteen}) second order in the $\pi$ fields),
corrections associated with the EM contact interactions
have not been taken into account.  In light of the
discussions above, we see that such treatments of EM
must, in fact, be inconsistent.

It is particularly easy to expose the problems of
such treatments in the case of $\pi\pi$ scattering.
To see this, note that
${\cal L}_{\pi ,EM}^{(0)}$
contains $\pi^4$ vertices and hence produces
contributions to $\pi\pi$ scattering amplitudes
zeroth order in the chiral expansion.  As noted
above, the radiative corrections are, in contrast,
second order, and hence expected to be considerably
smaller (as will be borne out by the results below).
Moreover, the set of radiative corrections do not
even exhaust the ${\cal O}(q^2)$ EM contributions
to $\pi\pi$ scattering; there exist one-loop graphs
involving vertices from
${\cal L}_{\pi ,EM}^{(0)}$
and tree graphs involving vertices from
${\cal L}_{\pi ,EM}^{(2)}$
(where the superscript is again the chiral order;
for the form of 
${\cal L}_{\pi ,EM}^{(2)}$
see Ref.~\onlinecite{urech}), all of which contain
no explicit photon but nonetheless produce EM contributions
at ${\cal O}(q^2)$.  It is thus first, inconsistent
to include only the ${\cal O}(q^0)$ photon exchange
contributions without including the ${\cal O}(q^0)$
contributions from
${\cal L}_{\pi ,EM}^{(0)}$
and, second, inconsistent to include the radiative
corrections without also including the other 
contributions, without explicit photons, of the
same chiral order.  

In this letter, we evaluate the 
\oqz
${\cal L}_{\pi ,EM}^{(0)}$
contributions to the s-wave $\pi\pi$ scattering 
lengths, and point out the problems that would ensue
from an incomplete ``EM subtraction'', {\it i.e.}
one involving only the \ogame contributions.  We
will return to the full set of \oq2 EM contributions
in a later publication\onlinecite{krmcwnew}.
Note that {\it no} version of the typical
meson-exchange-model treatment involving only graphs
with explicit photons would incorporate
the effects of the \oqz EM contact terms.  It is
usually said that the form factors employed in
such models are a phenomenological means of incorporating
short-distance effects.  However, such form factors,
in the low-energy effective theory, are generated by
higher-chiral-order corrections (loops, as well as
tree graphs involving higher order vertices) to the
leading tree-level vertices.  Hence, in the case of
the EM interactions, incorporating such form factors
in graphs with explicit photons produces only higher
order corrections, which begin at \oq2 and hence
could not possibly account for the contributions
associated with the \oqz contact terms.

We now turn to the evaluation of the non-\ogame
\oqz EM contributions to $\pi\pi$ scattering.  In
order to optimize our numerical accuracy, we will work
with the $SU(2)\times SU(2)$ effective Lagrangian,
involving the pions alone, the $SU(2)\times SU(2)$
chiral symmetry being much better respected than
$SU(3)\times SU(3)$.  The form of
${\cal L}_{\pi ,EM}^{(0)}$
is the same as in Eqn.~(\ref{fourteen}) except that
$U$ is now given by $U=\exp ({\rm i}\vec{\tau}\cdot\vec{\pi} /F)$,
$Q$ is the $u-d$ quark charge matrix, and we have
three, rather than eight, pseudoscalar fields.  The
part of 
${\cal L}_{\pi ,EM}^{(0)}$
second order in the fields is then easily shown to be
$-2c_\pi \pi^+\pi^-$
which produces the well-known non-zero EM contribution
to the charged pion mass, $(\delta m^2_{\pi^\pm} )_{EM}=2c_\pi$.
Since, as is also well-known\onlinecite{gl82}, the
$\pi^\pm$-$\pi^0$ mass splitting is essentially
purely EM, this fixes the value of $c_\pi$:  
$c_\pi = (m^2_{\pi^+}-m^2_{\pi^0})/2$.  Expanding, similarly,
to fourth order in the fields, one finds a contribution
\begin{equation}
{\frac{c_\pi}{3F^2}}\left[ 2{\pi^0}^2\pi^+\pi^-\ +\ 
4{\pi^+}^2{\pi^-}^2\right]\ .\label{nineteen}
\end{equation}
With the definitions
\begin{equation}
T(s,t)\ =\ 32\pi \sum_{\ell =0}^\infty (2\ell +1)
P_\ell \left( cos(\theta_{CM})\right) t_\ell (s)\label{twenty}
\end{equation}
for the partial wave amplitudes, and
\begin{equation}
{\rm Re}\left( t_0(s)\right) \ =\ a_0\ +\ {\cal O}(q_{CM}^2)
\label{twentyone}
\end{equation}
for the s-wave scattering lengths, where $\theta_{CM}$,
$q_{CM}$ are the scattering angle and magnitude of
the three-momentum in the CM frame, one obtains the
following $1\gamma E$-subtracted s-wave scattering lengths
at leading order in the EM chiral expansion
\begin{eqnarray}
&&a_0(00;00)\ =\ {\frac{1}{3}}a_0^0+{\frac{2}{3}}a^2_0
\nonumber \\
&&a_0(+0;+0)\ =\ {\frac{1}{2}}a^2_0+{\frac{c_\pi}
{24\pi F^2}}\nonumber \\
&&a_0(+-;00)\ =\ {\frac{1}{3}}a^0_0-{\frac{1}{3}}a^2_0
+{\frac{c_\pi}{24\pi F^2}}\nonumber \\
&&a_0(+-;+-)\ =\  {\frac{1}{3}}a^0_0+{\frac{1}{6}}a^2_0
+{\frac{c_\pi}{6\pi F^2}}\nonumber \\
&&a_0(++;++)\ =\ a^2_0+{\frac{c\pi}{6\pi F^2}}
\label{twentytwo}\end{eqnarray}
where $a^{0,2}_0$ are the strong $I=0,2$ s-wave
scattering lengths and
the arguments of $a^{0,2}_0$ refer to the initial and final
charge states (e.g., $(00;00)$ means the process
$\pi^0\pi^0\rightarrow\pi^0\pi^0$).
At lowest chiral order, $a_0^0 = 7m_\pi^2/32\pi F^2=0.16$
and $a^2_0 = -m_\pi^2/16\pi F^2=-0.045$, the Weinberg
values for the s-wave $I=0,2$ scattering lengths.  The
${\cal O}(q^4)$ strong corrections to these relations were 
worked out long ago by Gasser and Leutwyler\onlinecite{gl84},
producing the usually quoted one-loop values
\begin{eqnarray}
\left( a^0_0\right)_{1-loop}\ =\ 0.20\pm 0.01\nonumber \\
\left( a^2_0\right)_{1-loop}\ =\ -0.042\pm 0.008\ \ .\label{twentythree}
\end{eqnarray}
Using these values, one finds that the EM contact corrections
range in size from $0$ (for $\pi^0\pi^0\rightarrow\pi^0\pi^0$)
to $10\%$ (for $\pi^+\pi^+\rightarrow\pi^+\pi^+$).  These
corrections, of course, also break the familiar isospin relations
between the various amplitudes.  We discuss the consequences
of this fact below.  Note that the $1\gamma E$-subtracted
radiative corrections to the scattering
lengths in Eqn.~(\ref{twentytwo}) are\onlinecite{roig76}
$0\%$, $-0.2\%$, $-0.6\%$, $-1.3\%$ and $+2.2\%$, respectively.
As expected from the chiral counting, these are much
smaller than the \oqz contact corrections.  We stress
again that one should not add the radiative corrections
of Ref.~\onlinecite{roig76} to those of Eqn.~(\ref{twentytwo})
since there are additional \oq2 EM corrections that 
have yet to be accounted for.

Let us now illustrate the dangers of ignoring the presence
of the contact interactions in \leffpunc .  We imagine
(though this is unlikely to be the case in practice any
time in the near future) having available experimental
data on all of the processes listed above, and having
performed a standard ``EM subtraction'', {\it i.e.}
having removed the \ogame contributions to the amplitude.
(Recall that, because of the Coulomb pole, one must
perform the \ogame extraction in order to even be able
to define the scattering lengths in the first place.)
If we (erroneously) assumed that this had removed all
EM effects, we could attempt to determine the isospin
breaking in the strong interaction contributions by
determining the discrepancies in the values of
$a^0_0$, $a_0^2$ extracted from various combinations
of the \ogame -subtracted amplitudes which would be
equal in the isospin limit.  (The analogue to the
usual proceedures used to extract strong interaction
charge-independence- and charge-symmetry-breaking
NN observables should be obvious.)  For example,
in the isospin limit,
\begin{equation}
a_0^2\ =\ 2\left[ a_0(00;00)-a_0(+-;+-)\right]
\ =\ 2\left[ a_0(+-;+-)-a_0(+-;00)\right]\ \ .\label{twentyfour}
\end{equation}
Using Eqns.~(\ref{twentytwo}),(\ref{twentythree}),
however, one would find
\begin{eqnarray}
&&2\left[ a_0(00;00)-a_0(+-;+-)\right]
\ =\ -0.0507\nonumber \\
&&2\left[ a_0(+-;+-)-a_0(+-;00)\right]\ =\ -0.0355\label{twentyfive}
\end{eqnarray}
a discrepancy of $43\%$.  Similarly,
\begin{eqnarray}
&&3\left[ a_0(00;00)-{\frac{2}{3}}a_0(++;++)\right] \ = \ 
0.191\nonumber \\
&&3\left[ a_0(+-;+-)-{\frac{1}{6}}a_0(++;++)\right]\ =\ 
0.211\label{twentysix}
\end{eqnarray}
(where both expressions would be equal to $a_0^0$ in
the isospin limit), an $11\%$ discrepancy.  One would
then conclude that there was very large strong 
({\it i.e.} generated by $m_d-m_u\not= 0$) isospin
breaking in $\pi\pi\rightarrow\pi\pi$, a conclusion
which, in fact, is completely erroneous, as noted
already above\onlinecite{gl84}.

It should be pointed out that the discrepancies displayed
in Eqns.~(\ref{twentyfive}), (\ref{twentysix}) 
are the largest possible, {\it i.e.}
that all other ``extractions'' of $a_0^0$, $a_0^2$ 
would lie in the ranges bracketed by the quoted values.
If we consider the processes actually used in the
latest experimental extractions ($\pi^+\pi^+\rightarrow
\pi^+\pi^+$ and $\pi^+\pi^-\rightarrow\pi^0\pi^0$,
obtained via an analysis of $\pi^+p\rightarrow
\pi^+\pi^+n$ and $\pi^- p\rightarrow\pi^0\pi^0 n$
near threshold\onlinecite{bkm95}), then the corrections
are smaller.  Note that the \ogame contributions are
very forward peaked and would not affect the experimental
analysis, whereas the contact interactions are purely
s-wave, and hence would.  If we subtract the contact
EM contaminations of \oqz , the corrections to
$a_0^0$, $a_0^2$ are $-0.0076$ and $-0.0043$, respectively,
lowering the central value of $a_0^0$ extracted
experimentally from $0.21$ to $0.20$ and that of
$a_0^2$ from $-0.031$ to $-0.035$ ($4\%$ and $14\%$
corrections, respectively), in both cases, in improved
agreement with the predictions of (strong) ChPT to one
loop.  Although these corrections are smaller than the
existing $\pm 0.07$, $\pm 0.007$ errors on the
experimental extractions, this is only marginally so for the
$a_0^2$ case, which is the more favorable of the two
for future improvements in the accuracy of the experimental
determination, as explained in Ref.~\onlinecite{bkm95}.
It should also be noted that there will be additional
isospin-breaking effects in $\pi N\rightarrow \pi\pi N$
which have not yet been considered in the analysis of
Ref.~\onlinecite{bkm95}, and which could further
alter the extracted values.  Nonetheless, one should
bear in mind that, without subtracting the known \oqz EM
contact contributions, the experimental values of
$a_0^0$ and $a_0^2$ extracted from $\pi^+\pi^+\rightarrow
\pi^+\pi^+$ and $\pi^+\pi^-\rightarrow\pi^0\pi^0$ would
be expected to be, not $0.20$ and $-0.042$, but
$0.21$ and $-0.038$, respectively.

In conclusion, we have considered the \oqz EM,
non-\ogame contributions to the $\pi\pi$ s-wave
scattering lengths.  The corrections are, in some cases,
quite large (a possibility noted previously by
Gasser\onlinecite{gasser95} and associated with the
smallness of the strong scattering lengths, which vanish in
the chiral limit).  We find that, accounting for these
corrections, the extracted values of $a_0^0$ and
$a_0^2$ are brought into closer agreement with the
results of ChPT to 1-loop.  Moreover, the calculation
demonstrates the unavoidable importance, in making
EM subtractions, of considering not only the \ogame
and radiative corrections, but also those contributions
associated with EM contact terms (involving no explicit
photons), which contact terms are {\bf necessarily}
present in any low-energy effective hadronic theory
representing the physics of the standard model.
The implications for evaluating EM corrections to 
other strong interaction observables (such as the
$\pi NN$ couplings and $NN$ scattering lengths) are
obvious.

\acknowledgments

KM would like to thank the T5 Group, Los Alamos National Labs, for its
hospitality during the course of this work, to acknowledge useful
conversations with Jim Friar, and to thank Ulf Meissner for clarifying
the present status of the extraction of the $\pi\pi$ scattering
lengths from the analysis of $\pi N\rightarrow\pi\pi N$.  The
continuing financial support of the Natural Sciences and Engineering 
Research Council of Canada is also gratefully acknowledged.

\end{document}